\documentclass[twocolumn,superscriptaddress]{revtex4-2}

\usepackage[english]{babel}
\usepackage[colorinlistoftodos, color=green!40, prependcaption]{todonotes}
\usepackage{amsthm}
\DeclareUnicodeCharacter{2212}{-}
\usepackage{mathtools}
\usepackage{physics}
\usepackage{xcolor}
\usepackage{graphicx}
\usepackage{comment}
\usepackage[left=23mm,right=13mm,top=35mm,columnsep=15pt]{geometry} 
\usepackage{lipsum}
\usepackage{csquotes}
\usepackage[pdftex, pdftitle={Article}, pdfauthor={Author}]{hyperref}
\usepackage{bm}
\usepackage{siunitx}
\usepackage{xr}
\usepackage{comment}

\addto\captionsenglish{}

\begin{document}


\title{Arbitrary mechanical memory encoding via nonlinear waves in bistable metamaterials
}

\author{Audrey A. Watkins}
\affiliation{John A. Paulson School of Engineering and Applied Sciences, Harvard University, Cambridge, MA 02138, USA}

\author{Giovanni Bordiga}
\affiliation{John A. Paulson School of Engineering and Applied Sciences, Harvard University, Cambridge, MA 02138, USA}

\author{Mingxing Mu}
\affiliation{School of Aerospace Engineering, Tsinghua University, Beijing 100084, China}

\author{Vincent Tournat}
\affiliation{Laboratoire d'Acoustique de l'Universit\'e du Mans, UMR 6613, Institut d'Acoustique -- Graduate School, CNRS, Le Mans Universit\'e, Le Mans, France}
\affiliation{John A. Paulson School of Engineering and Applied Sciences, Harvard University, Cambridge, MA 02138, USA}

\author{Katia Bertoldi}
\affiliation{John A. Paulson School of Engineering and Applied Sciences, Harvard University, Cambridge, MA 02138, USA}




\date{\today}

\begin{abstract}
Mechanical metamaterials composed of bistable elements have recently emerged as promising platforms for mechanical memory. Traditional approaches to writing information in these systems typically rely on localized actuation or predefined coupling schemes, which are often labor-intensive or lack adaptability. In this work, we introduce a one-dimensional metamaterial consisting of mass-in-mass bistable units that are statically decoupled yet dynamically switchable, allowing arbitrary mechanical information to be encoded through nonlinear waves applied at the boundary of the system. Through a combination of experiments and simulations, we demonstrate that tailored input signals can selectively trigger state transitions deep within the structure, enabling remote and programmable bit writing. This approach opens a new avenue for mechanical memory, harnessing the robustness of bistable elements and the tunability of nonlinear wave-driven actuation.
\end{abstract}

\maketitle

Mechanical metamaterials -- engineered materials with unconventional, tunable mechanical properties \cite{bertoldi2017flexible, jiao2023mechanical, barchiesi2019mechanical} -- have emerged as a promising avenue for achieving advanced capabilities in response to external stimuli. These  encompass energy trapping~\cite{shan2015multistable}, wave guiding and filtering~\cite{deymier2013acoustic,DORN2023102091,CelliGonella}, and shape morphing~\cite{dudek2025shapemorphingmetamaterials,jin2020kirigami,Ni2022}. Furthermore, the introduction of multistable building blocks capable of supporting multiple stable configurations has broadened the range of functionalities to include stable propagation of pulses over arbitrarily long distances in dissipative media~\cite{raney2016stable, Jin_2020}, switchable mechanical properties~\cite{silverberg2014using, bilal2017bistable,Chen_2021,Novelino_pnas}, and reconfigurable shapes~\cite{Jin_2020, haghpanah2016multistable}. Additionally, bistable elements have proven well-suited for mechanical memory applications, as their discrete stable states function similarly to digital bits ~\cite{Chen_2021,pechac2023mechanical,treml2018origami,liu2023cellular}. This has opened avenues for mechanical metamaterials capable of supporting memory storage \cite{masana2020origami, jules2022delicate, yasuda2017origami, pechac2023mechanical} and functioning as logic gates \cite{mei2021mechanical, treml2018origami, jiang2019bifurcation, raney2016stable,Novelino_pnas} and counters \cite{kwakernaak2023counting}. For all these reasons related to their ability to retain memory and be sensitive to their environment, multistable metamaterials offer the potential to play an important role in the development of embedded mechanical intelligence, smart materials, and mechanical computing~\cite{kwakernaak2023counting, yasuda2021mechanical, liu2023cellular, gutierrezprieto2025dynamicdrivingenablesindependent}. Although the memory contained in mechanical metamaterials does not rival the memory of today's computers in many respects, it does have a few niche advantages, such as frugality in terms of energy expenditure and robustness against intense electromagnetic fields~\cite{merkle1993two, merkle2018mechanical, mei2023memory}.

The realization of mechanical materials with memory requires the ability to write the state of each element. This has been successfully demonstrated through the use of decoupled bistable units, which can be reconfigured by applying localized inputs  \cite{Chen_2021}. However, this writing process is labor-intensive, as it involves addressing each unit separately. To simplify the process, researchers have shown that certain predetermined sequences can be programmed into the material structure by engineering the couplings between bistable elements~\cite{jules2022delicate,Liu2024}. Additionally, a dynamic control strategy has recently been proposed to enable transitions between arbitrary states via global rotational driving cycles ~\cite{gutierrezprieto2025dynamicdrivingenablesindependent}. Finally, transition wavefronts have been demonstrated as an effective mechanism for sequentially switching units from one state to another, similarly to a domino effect \cite{raney2016stable, pechac2023mechanical,NadkarniPRL}. 
\begin{figure*}[t!]
\begin{center}
\includegraphics[width=0.95\linewidth]{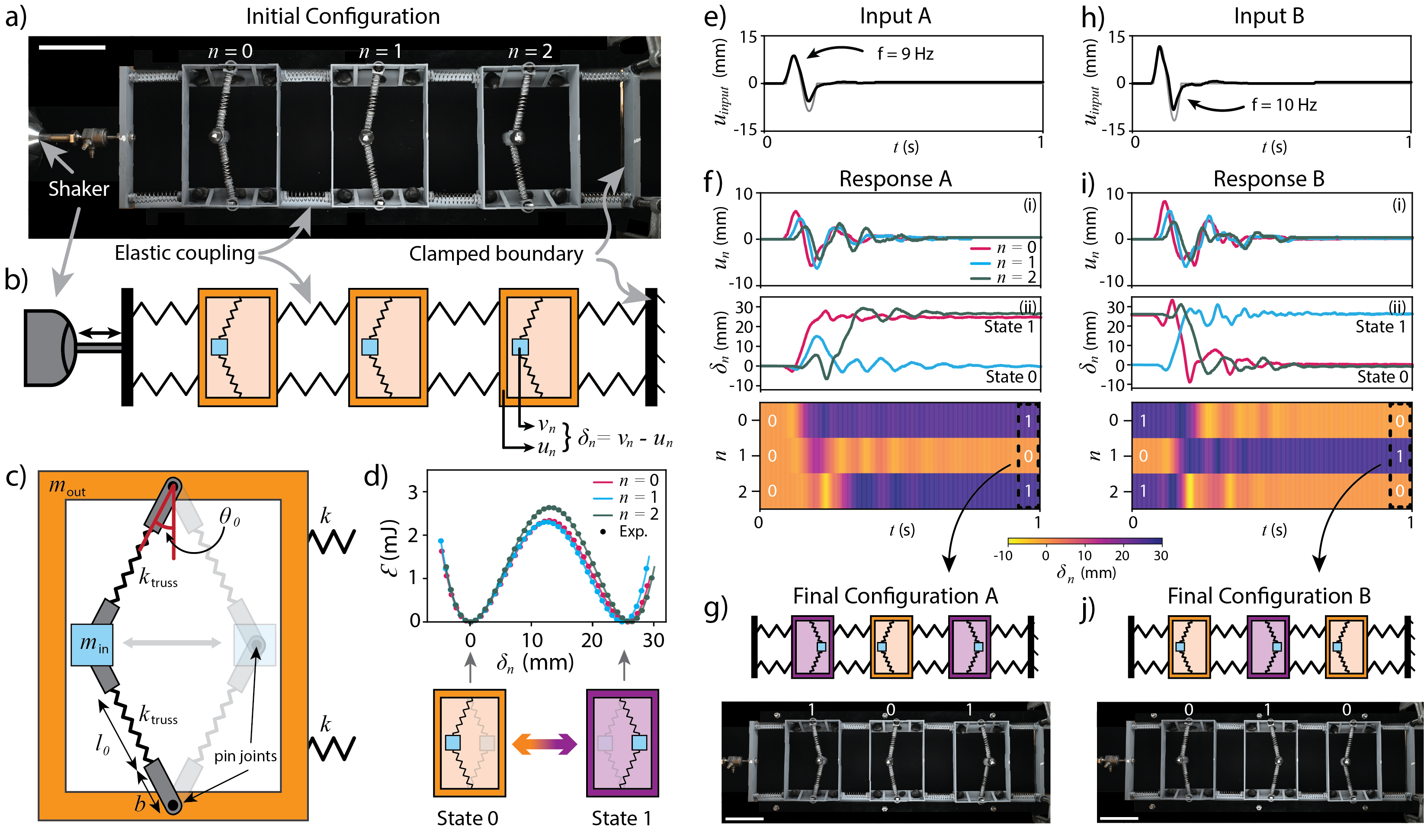}
\caption{\label{fig:bistable mechanical metamaterial} \textbf{Bistable mass-in-mass
metamaterial.} a) Photo and b) schematic of the 3-units bistable mass-in-mass metamaterial. c) Schematic of the unit cell. d) Experimentally characterized bistable elastic energy for the three von Mises trusses. e) Displacement input A (a bipolar pulse of amplitude $A=$8.8 mm and nominal frequency $f$=9 Hz) delivered by the shaker. f) Time and spatio-temporal response triggered by input A.  g) Final configuration triggered by input A. h) Displacement input B (amplitude $A=$11.8 mm and nominal frequency $f$=10 Hz), i) associated response, j) and final configuration. Scale bars: 5 cm.}
\end{center}
\end{figure*}

Here, we introduce a novel strategy for encoding arbitrary mechanical memory by harnessing nonlinear waves in a metamaterial made up of bistable units featuring a unique coupling architecture. Specifically, we employ a one-dimensional array of mass-in-mass units, each possessing two stable states that are statically decoupled yet dynamically switchable. This configuration enables the selective reconfiguration of individual bistable elements at arbitrary spatial positions within the structure.

Through a combination of experiments and simulations, we demonstrate that carefully tailored dynamic input profiles applied at one boundary can trigger controlled state transitions deep within the material, effectively writing mechanical bits. Additionally, we investigate how the distribution of mass within the unit cell influences the dynamic response of the system, allowing us to reduce unwanted sensitivity and enhance spatial precision in actuation. This wave-driven mechanism offers a new approach to writing and storing information mechanically, by harnessing dynamic reconfiguration of bistable elements.

Our mechanical metamaterial consists of a one-dimensional array of $N$ elastically coupled unit cells, each containing a bistable inclusion (Fig.~\ref{fig:bistable mechanical metamaterial}a). Each unit cell comprises two masses: an outer mass $m_{\text{out}}$ and an inner mass $m_{\text{in}}$. Adjacent outer masses are connected via two linear springs in parallel each with stiffness $k$, allowing mechanical waves to propagate through the structure. Additionally, each outer mass is connected to its corresponding inner mass through a bistable von Mises truss, which consists of two linear springs, each with stiffness $k_{\text{truss}}$ and rest length $l_{\text{0}}$ forming an angle $\theta_0$ to the vertical direction in the undeformed configuration. One end of each spring is connected to the outer mass via near-frictionless pin joints of fixed length $b=16$ mm, while the other end is connected to the inner mass $m_{\text{in}}$ (Fig.~\ref{fig:bistable mechanical metamaterial}c). Let $\delta_n = v_n - u_n$ denote the relative horizontal displacement between the inner and outer masses for the $n$-th unit cell, where $v_n$ and $u_n$ represent their respective displacements (Fig.~\ref{fig:bistable mechanical metamaterial}b). In this framework, stable \emph{state 0} corresponds to $\delta_n=\delta^{(0)} = 0$, while \emph{state 1} occurs at $\delta_n=\delta^{(1)} = 2l_0 \tan \theta_0$ (Fig.~\ref{fig:bistable mechanical metamaterial}d). 

Under quasi-static axial loading, this mass-in-mass architecture behaves like a typical mass-spring system. The applied load cannot switch the state of the individual von Mises trusses, and the configuration of these trusses does not influence the overall system response  (see Video S1).  In contrast, when a dynamic pulse is applied at the boundary of the metamaterial, both $m_{\text{in}}$ and $m_{\text{out}}$ begin to oscillate due to inertial coupling, which links their motions. This interdependence between the two degrees of freedom enables the encoding of arbitrary mechanical memory via nonlinear waves excited at the boundaries.
To illustrate this, in Fig. \ref{fig:bistable mechanical metamaterial}, we focus on a metamaterial with $N=3$ unit cells, $m_{\text{in}}=30$ grams, $m_{\text{out}}=70$ grams, $k=125$ N/m, $k_{\text{truss}}=1428 \pm 72$ N/m, $l_{\text{0}}=28.8 \pm 1$ mm and $\theta_0=11.55 \pm 0.14^{\circ}$, where the slight variations in the parameters defining the von Mises trusses arise from minute fabrication imperfections.  In our experiments, one end of the metamaterial is fixed, while the other is connected to a low-frequency shaker that delivers a bipolar pulse (generated by providing a single period of a sinusoidal electrical signal), characterized by an amplitude $A$ and frequency $f$. The response of the metamaterial is recorded using an overhead high-speed camera, and a point-tracking algorithm is employed to extract the displacements of the input ($u_{\text{input}}$) as well as the inner and outer masses of the $n$-th unit ($v_n$ and $u_n$, respectively). Using this data, we compute the relative displacement for each unit, $\delta_n(t)$, and identify the updated state of the metamaterial -- defined by the set of unit states $\alpha_k = (\beta_1,\beta_2,\ldots,\beta_N)\in\{0,1\}$, with $k=0,1,\ldots,2^N-1$. In this study, the $k$-th state corresponds to the binary representation of the integer $k$  (i.e., for a system with $N=3$,  $\alpha_0 = (0,0,0)$, $\alpha_1 = (0,0,1)$, $\alpha_2 = (0,1,0)$, etc.). 

Three key features emerge from the results shown in Fig.~\ref{fig:bistable mechanical metamaterial}. First, as illustrated by $u_n(t)$ in Fig.~\ref{fig:bistable mechanical metamaterial}f, a pulse with $A=8.8$ mm and $f=9$ Hz initiates a wave that propagates through the structure, reflects multiple times, and eventually dissipates. Second, when the accelerations of the outer masses are sufficiently large, the resulting inertial forces on the inner masses can overcome their energy barriers, causing them to switch states. This is evident from the evolution of $\delta_n(t)$, which transitions to $\delta^{(1)}$ for units 0 and 2. As such, after the wave dies out, the metamaterial settles into the new state $\alpha_5 = (1,0,1)$ (Figs.~\ref{fig:bistable mechanical metamaterial}f and~\ref{fig:bistable mechanical metamaterial}g).
Third, by applying a series of sufficiently strong pulses, we can reprogram the state of the metamaterial in a robust and repeatable manner. For example, starting from the updated state $\alpha_5 = (1,0,1)$, the application of another pulse with $A=11.8$ mm and $f=10$ Hz drives the system to transition to the state $\alpha_2 = (0,1,0)$  (Fig.~\ref{fig:bistable mechanical metamaterial}i and~\ref{fig:bistable mechanical metamaterial}j).
\begin{figure}
\centering
\includegraphics[width=0.95\linewidth]{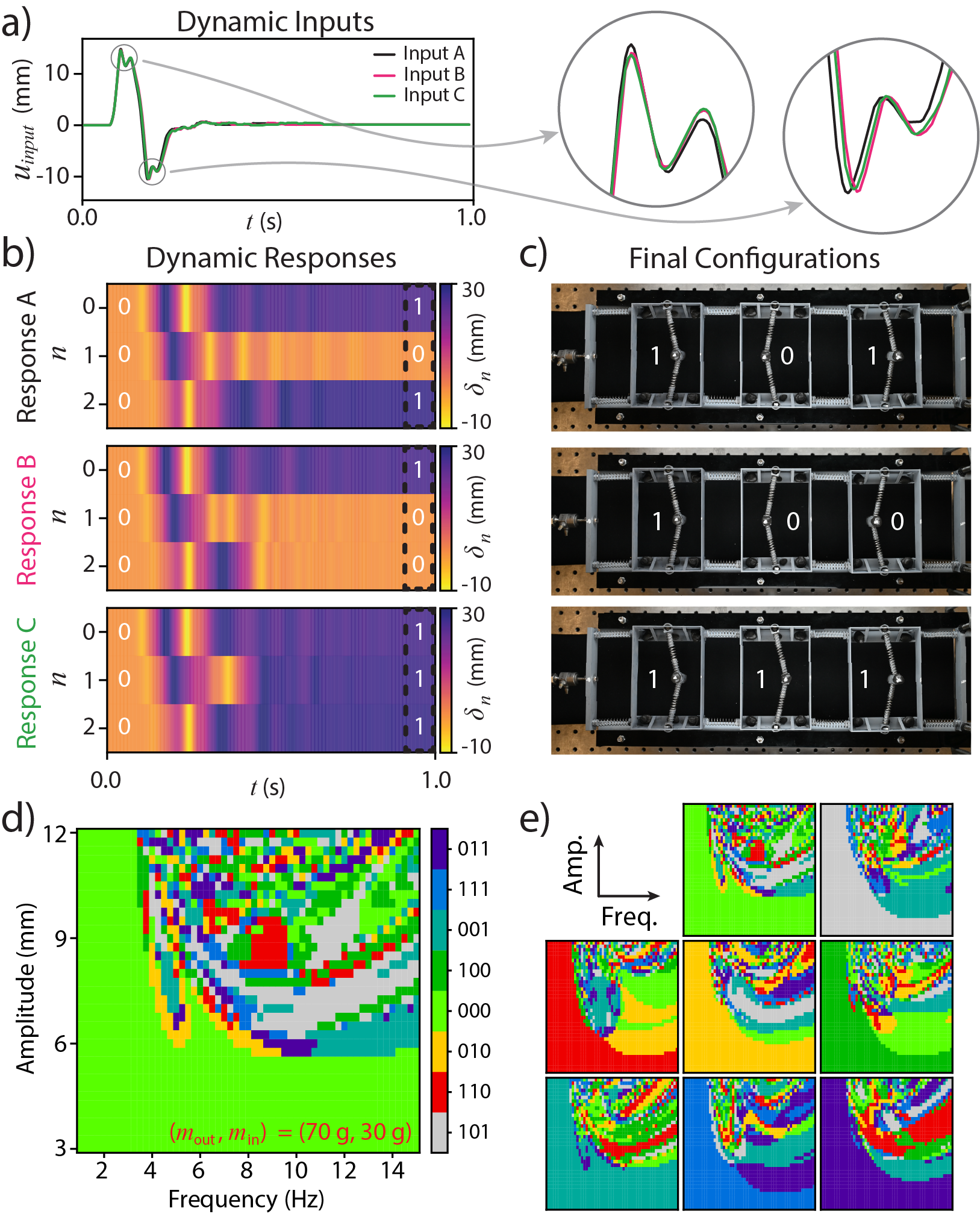}
\caption{\textbf{Effect of small perturbations in the input signal.} \small{a) 3 slightly different input signals, and b) spatio-temporal plots of the corresponding responses, with c) photos of the 3 different final states reached. d)-e) Numerically predicted  final states as a function of input frequency and amplitude for a sinusoidal input  reached from d) $\alpha_0 = (0,0,0)$ and e) the seven other states.}}
\label{fig:sensitivity analysis}
\end{figure}

The results in Fig.~\ref{fig:bistable mechanical metamaterial} demonstrate that sequences of sufficiently large-amplitude pulses applied at the boundary of the metamaterial can be used to encode arbitrary internal states. However, the final state imprinted by the wave is highly sensitive to small perturbations in the input signal. To illustrate this sensitivity, in Fig.~\ref{fig:sensitivity analysis} we consider the same system initialized in the $\alpha_0 = (0,0,0)$ state, subjected to three pulses. These pulses are nearly identical with $(A, f) = (13.5~\text{mm},\, 7.2~\text{Hz}), (13.3~\text{mm},\, 6.9~\text{Hz})$, and $(13.2~\text{mm},\, 7.0~\text{Hz})$ (Fig.~\ref{fig:sensitivity analysis}a). Despite the minimal differences in the input signals, they drive the metamaterial into three distinct final states: $\alpha_5 = (1,0,1)$, $\alpha_4 = (1,0,0)$, and $\alpha_7 = (1,1,1)$ (Fig.~\ref{fig:sensitivity analysis}b).

To understand the sensitivity of the metamaterial to the applied input and to investigate whether less sensitive configurations exist, we develop a model. Towards this end, we write the equations of motion for the $n$-th unit cell as
\begin{equation}
\begin{aligned}
m_{\text{out}} \ddot{u}_n + k (2u_n - u_{n+1} - u_{n-1}) + &c_{\text{out}} \dot{u}_n \\ - c_{in}\dot{\delta}_n + 
        f_{fr}(\dot{u}_n&)- \mathcal{E}'(\delta_n) =0,  \\
m_{\text{in}} \ddot{v}_n+ c_{\text{in}} \dot{\delta}_n +\mathcal{E}'(\delta_n)=0,~~~~~~~~~&
\end{aligned}
\label{eqn: motion equations}
\end{equation}
where $\ddot{(\cdot)}=\partial^2(\cdot)/\partial t^2$, $\dot{(\cdot)}=\partial(\cdot)/\partial t$, ${(\cdot)}'=\partial(\cdot)/\partial \delta_i$ and $\mathcal{E}$ is the potential of each bistable von Mises truss
\begin{equation}
   \mathcal{E}(\delta_n) =  k_{\text{truss}} \left( L \sec\theta_0-\sqrt{(L\tan\theta_0 - \delta_n)^2+L^2}\right)^2,
\end{equation}
with $L = l_{0}\cos\theta_0+2b$.
Furthermore, $f_{fr}(\dot{u}_n) = m_{\text{out}}g\mu_k \tanh (\beta \dot{u}_n)$ is the friction force between the outer mass and the platform supporting the structure with $g = 9.81$ ms$^{-2}$, $\beta = 100$ m$^{-1}$s, $\mu_k=0.0675$ is the kinetic coefficient of friction, and $c_{\text{out}}=0.22$ kg/s and $c_{\text{in}}=0.1$ kg/s are the viscous damping coefficients which are fit to experimental data (see Fig. S3). To assess the accuracy of our model, we numerically integrate Eq.~(\ref{eqn: motion equations}) using the Runge-Kutta 45 method with an adaptive step size, and compare the resulting predictions with our experimental data. In the simulations, we apply the experimentally measured input displacement $u_{\text{input}}$ to the leftmost boundary of a three-cell chain, while imposing zero-displacement boundary conditions at the right end. As shown in Fig. S4, we find agreement between experimental and numerical results, thereby validating the predictive capability of our model.

Having verified the accuracy of our model, we leverage it to systematically examine the sensitivity of the metamaterial to variations in the input signal. Specifically, we consider the metamaterial initially in each of the eight supported states and simulate its response to a single period of a sinusoidal input with $f \in [1, 15]$ Hz and $A \in [3, 12]$ mm. In Fig.~\ref{fig:sensitivity analysis}d, we focus on the case where the metamaterial begins in state $\alpha_0 = (0,0,0)$ and represent its final state as a colored pixel, plotted as a function of input frequency and amplitude. As expected, a prominent region of light green pixels appears in the plot, indicating the system remains in the state $\alpha_0 = (0,0,0)$ and confirming that low-frequency, low-amplitude inputs do not induce state transitions due to insufficient acceleration. In contrast, sufficiently high frequencies and amplitudes result in transitions to other states. However, consistent with experimental observations, the final state is highly sensitive to the specific input parameters, producing a sharply pixelated pattern. To quantify this sensitivity, we compute the largest contiguous pixel area associated with each final state $k$, denoted as $A_{\alpha_0}^{\alpha_k}$. Excluding the initial state ($k\neq0$), we find the maximum contiguous area to be $A_{\alpha_0}^{\alpha_5} = 123$ pixels$^2$ for state $\alpha_5 = (1,0,1)$, and the minimum to be $A_{\alpha_0}^{\alpha_7} = 10$ pixels$^2$ for state $\alpha_7 = (1,1,1)$. Finally, in Fig.~\ref{fig:sensitivity analysis}e, we extend this analysis to the remaining seven initial configurations. In all cases, the highly pixelated patterns persist, underscoring the strong dependence of the metamaterial’s response on the specific characteristics of the input signal. 
\begin{figure}[t!]
\centering
\includegraphics[width=0.95\linewidth]{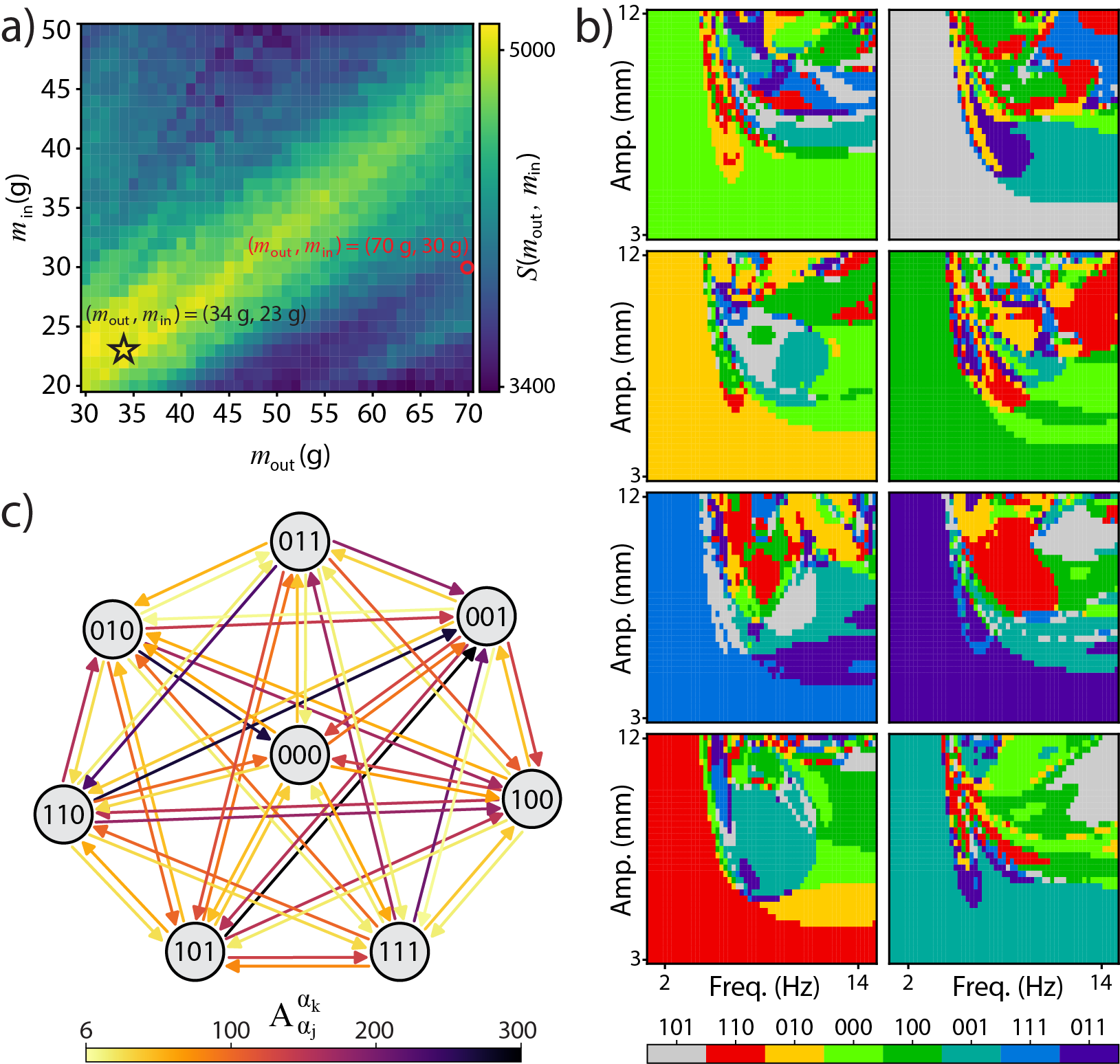}
\caption{\textbf{Mass-dependent response space.} a) Sensitivity function to input perturbations as defined by Eq.~(\ref{eqn: area obj function}) as a function of $m_{\text{out}}$ and $m_{\text{in}}$, with both configurations of masses used in this article. b) Color-coded final states reached from all possible initial configurations in response to a single period of sinusoidal input with a given amplitude and frequency (pixel size: $0.25$ mm $\times$ $0.25$ Hz). c) Transition graph with arrows between states indicating an accessible phase transition and color representing the corresponding maximum contiguous area size.}
\label{fig:simulated optimal mass response}
\end{figure}

Next, we investigate how variations in $m_{in}$ and $m_{out}$ affect the sensitivity of the metamaterial to the input signal. Towards this end, we vary $m_{in}\in[20,50]$ grams and $m_{out}\in [30, 70]$ grams and simulate the response of the metamaterial starting from each of the eight supported states. To summarize the results, in Fig. \ref{fig:simulated optimal mass response}a we report the evolution of 
\begin{equation}
    S = \sum^{2^n-1}_{j=0} \sum^{2^n-1}_{k=0,k\neq j} A^{\alpha_k}_{\alpha_j}(m_{\text{out}}, m_{\text{in}})
\label{eqn: area obj function}
\end{equation}
as a function of $m_{\text{in}}$ and $m_{\text{out}}$. Large values of $S$ indicate large contiguous regions and low pixelation across all eight initial configurations, while small values correspond to highly pixelated, sensitive responses. We find that, while the metamaterial considered in Fig.~\ref{fig:sensitivity analysis} yields $S = 3899$ pixels$^2$ with mass values $(m_{\text{out}}, m_{\text{in}}) = (70~\text{g}, 30~\text{g})$, the value of $S$ significantly increases with lower values of $m_{\text{out}}$ and $m_{\text{in}}$. For instance, $(m_{\text{out}}, m_{\text{in}}) = (34~\text{g}, 23~\text{g})$ yields $S = 5095$ pixels$^2$. The resulting final states for this configuration, as a function of input frequency and amplitude, are shown in Fig.~\ref{fig:simulated optimal mass response}b for each of the eight initial states. Overall, we observe a substantial reduction in pixelation across all cases, indicating reduced sensitivity to the input signal within the considered amplitude and frequency ranges.
Next, in Fig.~\ref{fig:simulated optimal mass response}c, we present the transition graph for this nearly-optimal metamaterial, with edges colored according to $A^{\alpha_k}_{\alpha_j}$. First, we observe that all eight states are dynamically accessible, regardless of the metamaterial’s initial configuration. Second, we note varying levels of sensitivity to the applied dynamic input among the transitions: some exhibit low sensitivity to the input, corresponding to darker arrows that reflect large contiguous regions in the parameter space; others show higher sensitivity and are represented by lighter arrows, which reflect smaller contiguous regions. 

\begin{figure}[h]
\centering
\includegraphics[width=0.95\linewidth]{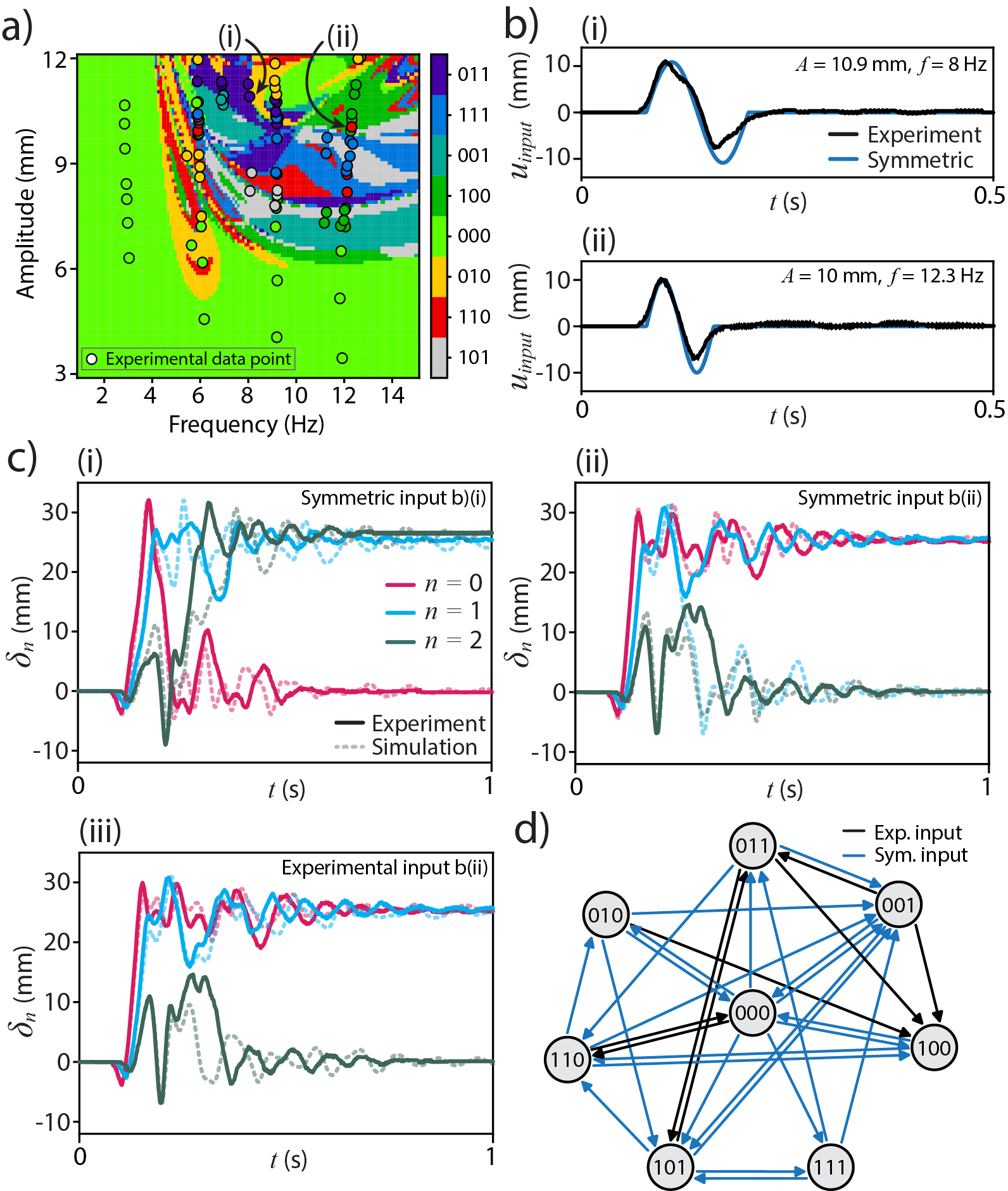}
\caption{\textbf{Experimental results.} a) Color-coded final state heat map (pixel size: $0.1$ mm $\times$ $0.1$ Hz) with experimental data points overlaid. b) Experimental input and fitted pulse with i) $(A, f) = (10.9~\text{mm},\, 8~\text{Hz})$ and ii) $(A, f) = (10~\text{mm},\, 12.3~\text{Hz})$. c) Experimental and numerically simulated $\delta_n(t)$ in response to c(i)) symmetric input $(A, f) = (10.9~\text{mm},\, 8~\text{Hz})$, c(ii)) symmetric input $(A, f) = (10~\text{mm},\, 12.3~\text{Hz})$, and c(iii)) experimental input $(A, f) = (10~\text{mm},\, 12.3~\text{Hz})$. d) Experimentally realized and numerically predicted transition graph with blue arrows indicating a symmetric fitted input and black arrows indicating the experimental input.}
\label{fig:exp writing}
\end{figure}

Guided by the results shown in Fig.~\ref{fig:simulated optimal mass response}, we modify the sample to have $m_{\text{in}} = 23$ grams and $m_{\text{out}} = 34$ grams, and systematically test its response. Starting from the initial state $\alpha_0 = (0,0,0)$, we apply bipolar pulses of varying frequencies and amplitudes. In Fig.~\ref{fig:exp writing}a, each experimental outcome is represented by a circular marker, colored according to the final state and overlaid on the corresponding numerical predictions. Overall, the experiments follow the trends predicted by the simulations, though some discrepancies arise.
These discrepancies primarily stem from distortions in the experimentally applied input signals, such as additional harmonics (Fig.~\ref{fig:exp writing}b(i)) and asymmetry between the positive and negative lobes (Fig.~\ref{fig:exp writing}b(i) and b(ii)). While such distortions do not always affect the final state, they can be significant enough to alter the  response of the system. For example, the input shown in Fig.~\ref{fig:exp writing}b(i) is best approximated by a symmetric sinusoidal pulse with ($A$, $f$) = (10.9 mm, 8 Hz). When excited with such a sinusoidal input, the model predicts a final state of (0,1,1), in agreement with the experiment (Fig.~\ref{fig:exp writing}c(i)). In contrast, the input in Fig.~\ref{fig:exp writing}b(ii) is best fit by a sinusoidal pulse with ($A$, $f$) = (10 mm, 12.3 Hz). For this sinusoidal input, the model predicts a final state of (1,0,0), whereas the experiment yields (1,1,0) (Fig.~\ref{fig:exp writing}c(ii)). However, when the actual experimental waveform from Fig.~\ref{fig:exp writing}b(ii) is used as input to the model, the simulated dynamics closely match the experimental outcome (Fig.~\ref{fig:exp writing}c(iii)).

Finally, we conducted experiments starting from each of the remaining seven initial states, using the plots in Fig.~\ref{fig:simulated optimal mass response}b to guide us in targeting all possible final states. The experimental outcomes are summarized in the transition graph shown in Fig.~\ref{fig:exp writing}d (see also Fig.~S5). Blue arrows indicate transitions observed experimentally that match those predicted numerically using the best-fitting sinusoidal input.
As shown, a substantial number of numerically predicted transitions are successfully realized in the experiments. However, 32 predicted transitions were not verified experimentally. Interestingly, eight of these “missing” transitions do match the numerical predictions when using the actual experimental input signals (black arrows in Fig.~\ref{fig:exp writing}d and Fig.~S8).
These findings highlight that, while the optimized metamaterial exhibits significantly improved robustness to input variations compared to the initial configuration, the precise waveform of the input remains a key factor in determining the system's response.
 
To summarize, we have demonstrated that nonlinear waves can be harnessed to encode arbitrary mechanical memory in a one-dimensional array of bistable mass-in-mass units coupled through linear springs. This strategy enables efficient information encoding, as a single boundary pulse is sufficient to set a desired state. A key challenge of the proposed approach lies in the sensitivity of the final state to the input signal, where small, unavoidable variations in the pulse can result in significantly different encoded states. However, we have shown that the system can be optimized to reduce its sensitivity to such variations. While our optimization focused solely on tuning the masses, further improvements can be achieved by tailoring the properties of the bistable oscillators (see SI Section IV). Although this study focused on a system with three units, the results readily extend to larger arrays containing more memory bits (see SI Section IV). Furthermore, generalizing this concept to two-dimensional tessellations could open new opportunities for information encoding and mechanical sensing.\\

\noindent
\emph{Acknowledgments}: This work was supported by the Simons Collaboration on Extreme Wave Phenomena Based on Symmetries, by the National Science Foundation (NSF) under award no. 2118201 and by CNRS via IRP DynaMetaFlex. \\

\noindent
\emph{Data availability}: 

The data that support the findings of this Letter are openly available at \href{https://github.com/bertoldi-collab/mass_in_mass}{github.com/bertoldi-collab/mass$\_$in$\_$mass}

\end{document}